\documentclass[twocolumn,twoside,slac_two]{revtex4}
\usepackage{graphicx}
\usepackage{fancyhdr}
\begin{document}

\title{CP Violation in the Neutrino Sector}


\author{Stephen J. Parke}
\affiliation{Theoretical Physics Department,
Fermi National Accelerator Laboratory \\
P.O.Box 500, Batavia, IL 60510, USA\\
parke@fnal.gov}

\begin{abstract}
This talk will address how various experiments will address the following issues: the $\nu_e$ flavor content of the 3rd neutrino mass eigenstates, $\sin^2 \theta_{13}$, the mass ordering of the neutrinos,
${\rm sign}(\delta m^2_{31})$ and whether CP is violated in the neutrino sector, $\sin \delta \neq 0$.
\end{abstract}

\maketitle

\thispagestyle{fancy}


\section{Introduction}
\begin{figure}[h]
\centering
\includegraphics[width=80mm]{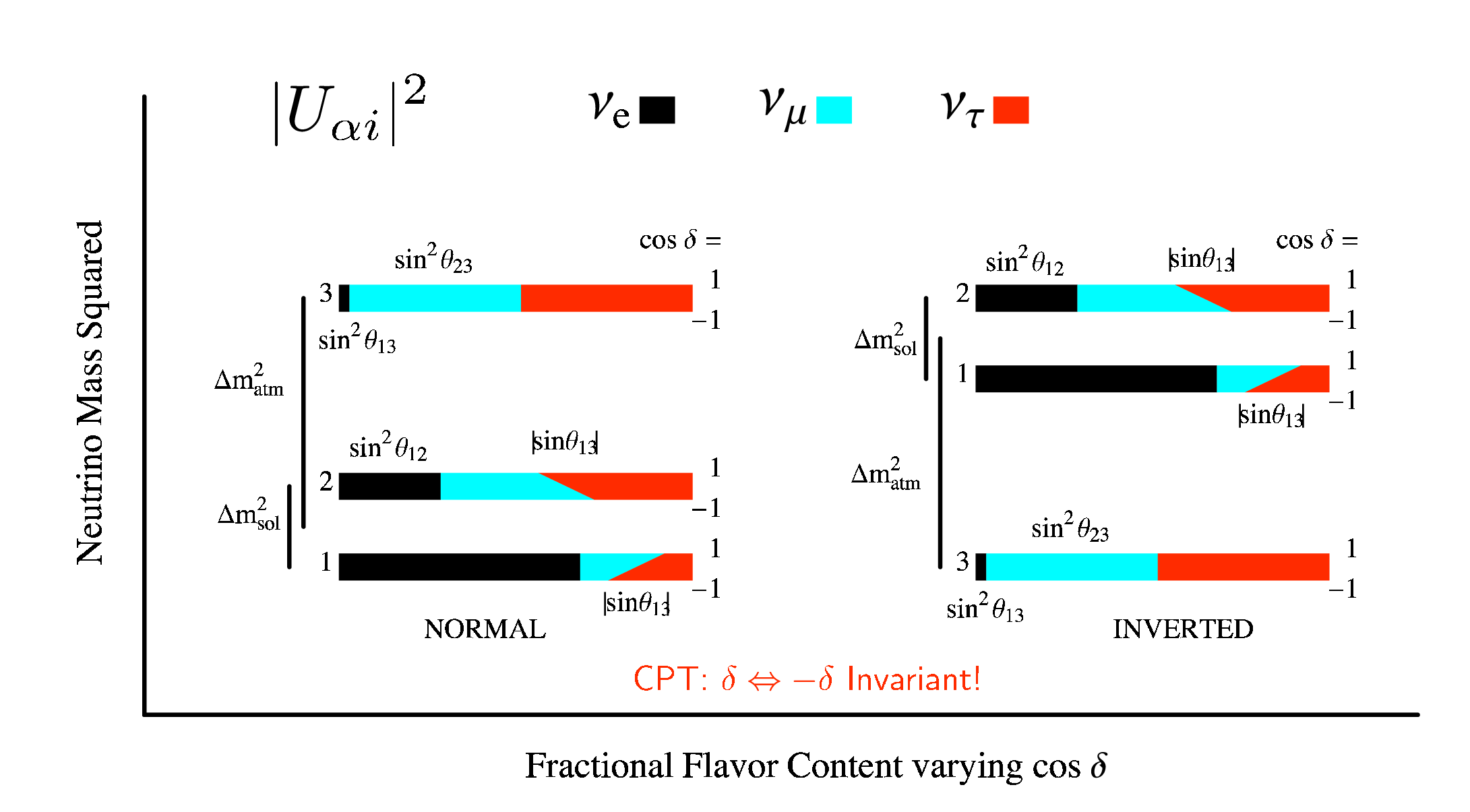}
\caption{Flavor fraction of the three neutrino mass eigenstates showing the dependence on the
cosine of the CP violating phase, $\delta$. If CPT is conserved, the flavor fraction must be the same
for neutrinos and anti-neutrinos. This figure was adapted from Ref.~\protect \cite{pmns-fig}.} \label{fig:pmns-sq}
\end{figure}

Fig.\ref{fig:pmns-sq} summarizes are current knowledge of the flavor content as well as the mass ordering of the neutrino mass eigenstates assuming that there are only three flavors of neutrinos. 

Our current knowledge of the $\delta m^2$'s can be summarized as follows
\begin{eqnarray}
\vert \delta m^2_{32} \vert & = & 2.4\pm 0.4 \times 10^{-3} {\rm eV^2} \nonumber \\ {\rm and}  \quad \quad
\delta m^2_{21} & = & + 7.6 \pm 0.4 \times 10^{-5} {\rm eV^2}
\end{eqnarray}
where the measurement of $\vert \delta m^2_{32} \vert$ comes from the MINOS experiment and that of
$\vert \delta m^2_{12} \vert$ from the KamLAND experiment, see \cite{Nu2008}.  The sign of  $\delta m^2_{21}$ is determined from the SNO experiment. 

The mixing angles and phase, using the particle data book convention, are given by
\begin{eqnarray}
\sin^2 \theta_{12} & = & 0.31\pm0.02  \nonumber \\
\sin^2\theta_{23} & = & 0.50\pm0.12  \nonumber \\
\sin^2 \theta_{13} & < & 0.04  \\
0 \le & \delta&  < 2 \pi \nonumber.
\end{eqnarray}
The best constraints on $\sin^2 \theta_{12} ,~\sin^2 \theta_{23} ~{\rm and} ~\sin^2 \theta_{13}$
come from SNO, SuperK's L/E analysis and Chooz respectively, see \cite{Nu2008}. Global fits make only marginal improvements on our knowledge of any of these parameters.

\section{The Unknowns}

The unknowns that can be addressed via neutrino oscillation experiments are
\begin{itemize}
\item The $\nu_e$ fraction in 3nd mass eigenstate:  $\sin^2\theta_{13}$
\item The neutrino mass hierarchy: $sign(\delta m^2_{32})$
\item Is CP violated: $\sin \delta \neq 0$
\item Is $|U_{\mu 3}|^2 <, =, > |U_{\tau 3}|^2$: $\sin ^2 \theta_{23} <, = ,> 1/2$
\item Unitarity of the MNS mixing matrix: \# of light sterile $\nu$'s
\item New Interactions and Surprises (the unknown unknowns)
\end{itemize}
The other important question is whether the light neutrinos are Majorana or Dirac which can be addressed in neutrinoless double beta decay.% and will not be further addressed here.

\section{{\Large \mathversion{bold}$\nu_e$} Disappearance}

\begin{figure}[b]
\centering
\includegraphics[width=80mm]{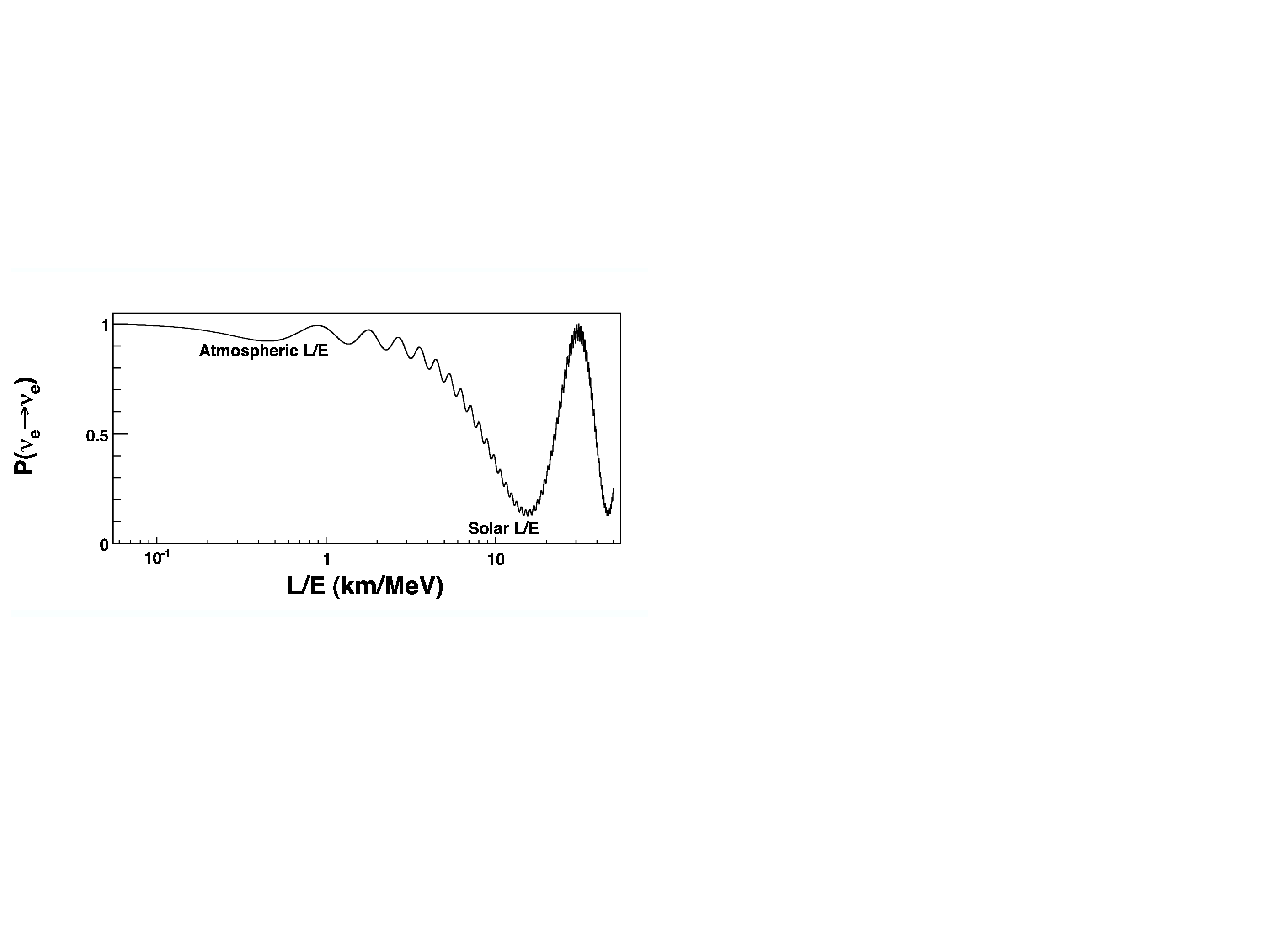}
\caption{The $\nu_e$ survival probability as a function of L/E showing both the atmospheric and solar oscillations..} \label{fig:nue_survival}
\end{figure}

\begin{figure*}[t]
\includegraphics[width=55mm]{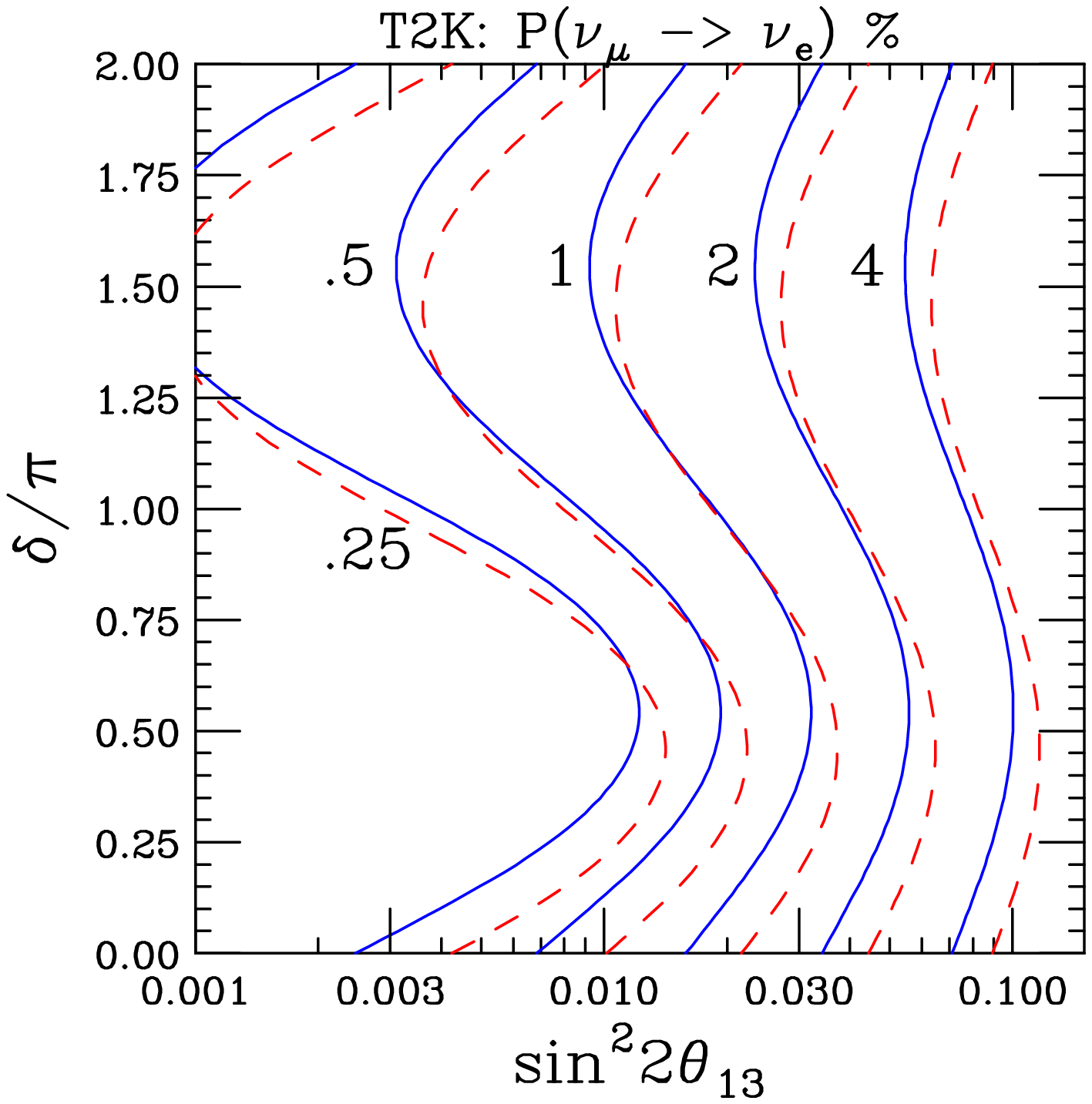}
\includegraphics[width=55mm]{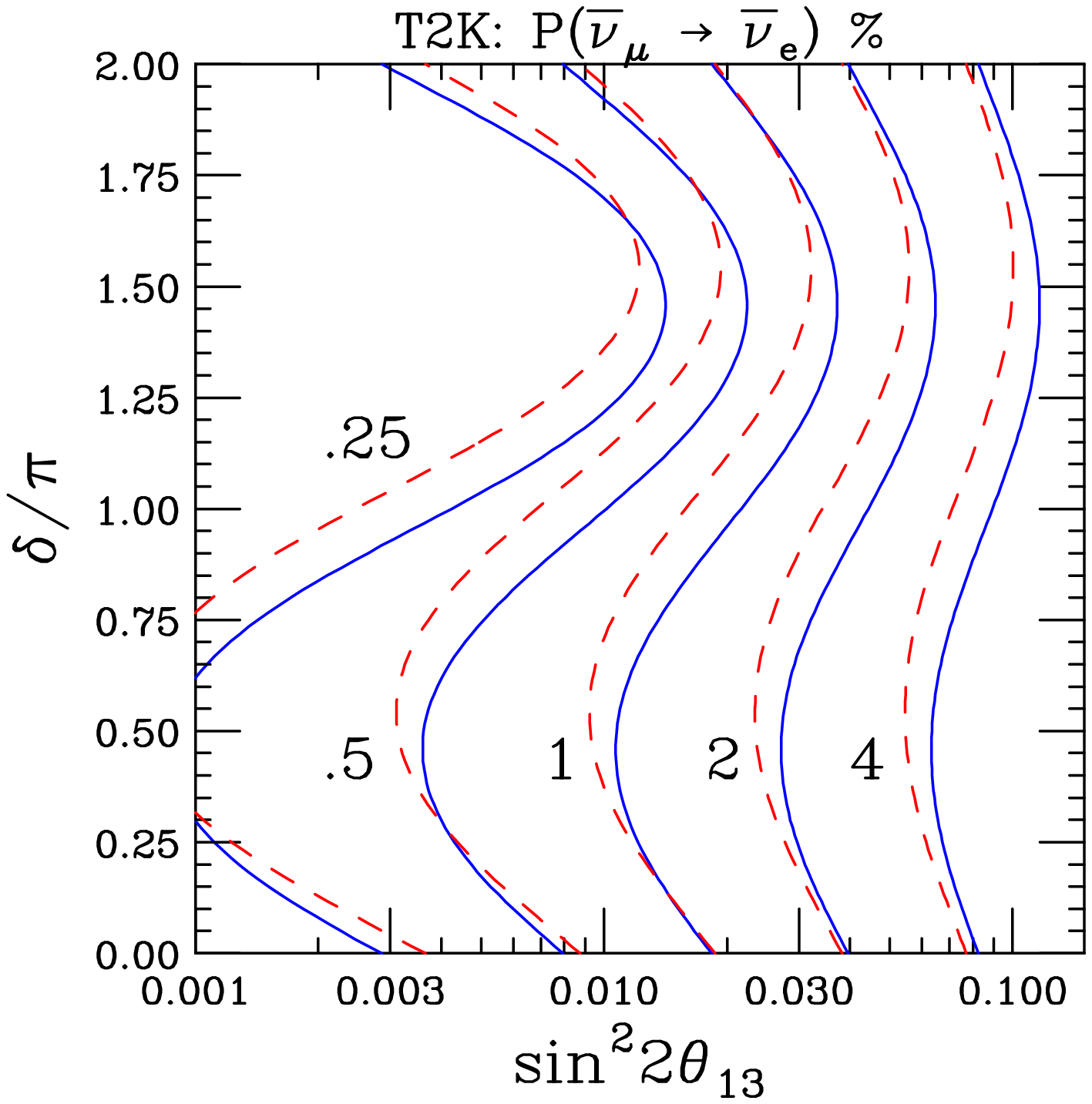}
\includegraphics[width=55mm]{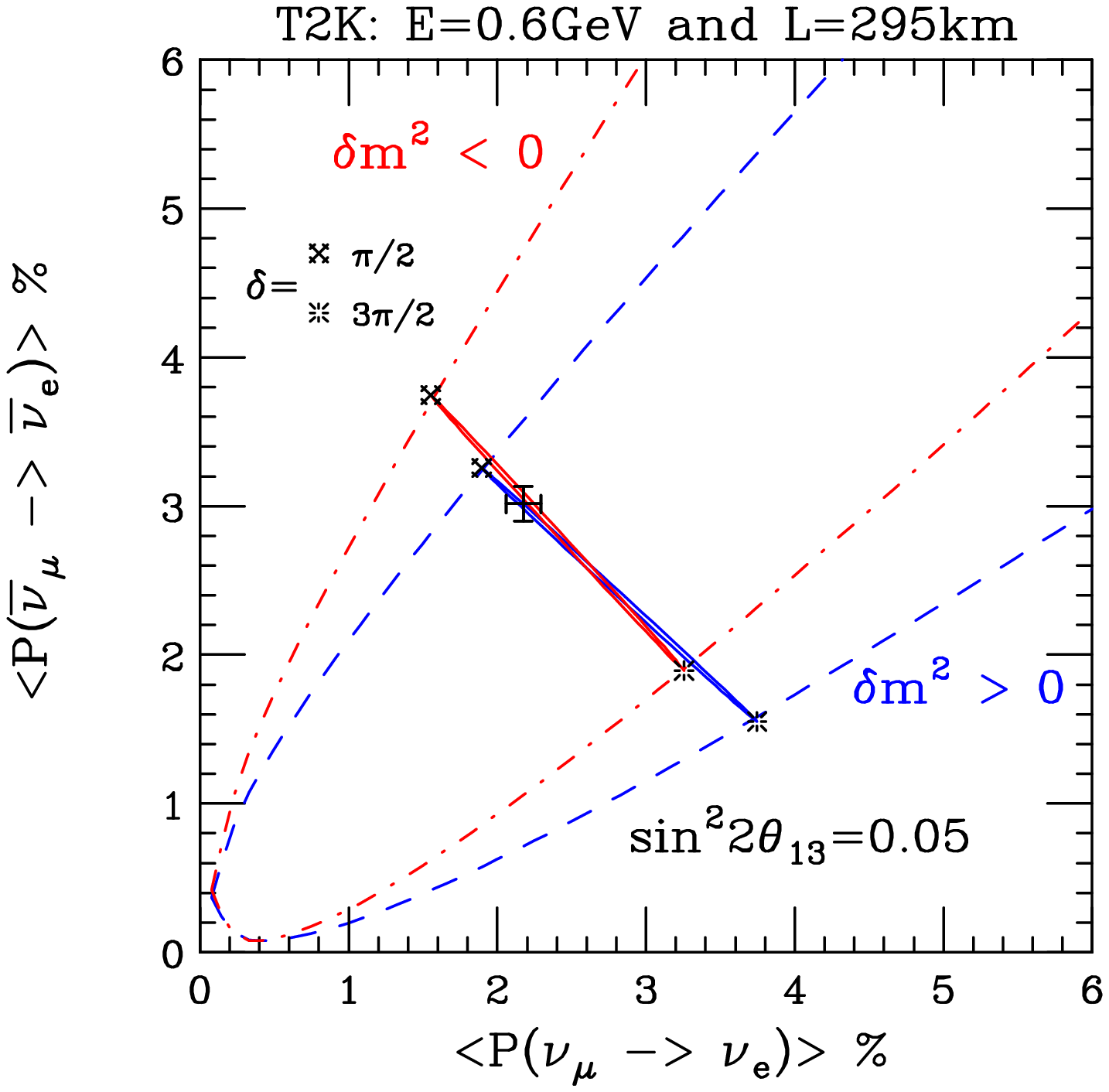}
%\vspace*{-1cm}
\caption{The left and middle panels are the iso-probability contours for T2K as a \% for the neutrino
(left) and anti-neutrino (middle) channels.  The solid (blue) line is for the normal hierarchy whereas the
dashed (red) line is for the inverted hierarchy.
The right panel is the bi-probability plot showing the correlation between the two probabilities.
The matter effect is small but non-negligible for T2K.}
\label{fig:t2kprobs}
\end{figure*}

The most direct way to address the $\nu_e$ fraction of the 3rd neutrino mass eigenstate is via reactor
neutrino disappearance experiments at the first atmospheric oscillation minimum, that is 1 to 2 km from the reactor core.  The $\nu_e$ survival probability in vacuum is given by (see Fig.\ref{fig:nue_survival})

\begin{eqnarray}
 & & P(\bar{\nu}_e \rightarrow \bar{\nu}_e) = 1 - \cos^4 \theta_{13}\sin^2 2\theta_{12} \sin^2 \Delta_{21}\nonumber \\ & &- \sin^2 2 \theta_{13} (\cos^2 \theta_{12} \sin^2 \Delta_{31} + \sin^2 \theta_{12} \sin^2 \Delta_{32}) \nonumber 
 \end{eqnarray}
 which can be rewritten as
 \begin{eqnarray}
 P(\bar{\nu}_e \rightarrow \bar{\nu}_e) & \approx &  1 - \sin^2 2 \theta_{13}  \sin^2 \left( \frac{\delta m^2_{ee} L}{4E}\right)   -{\cal O}(\Delta_{21}) ^2.  \nonumber
 \end{eqnarray}
Where $\Delta_{jk}$ is used as a shorthand for the the kinematic phase, $\delta m^2_{jk}L/4E$
and
\begin{eqnarray}
\delta m^2_{ee} & =  & \cos^2 \theta_{12} |\delta m^2_{31}|+ \sin^2 \theta_{12} | \delta m^2_{32}|
\end{eqnarray}
is the atmospheric $\delta m^2$ for the $\nu_e$ survival probability. This is the electron flavor weighted average of $|\delta m^2_{31}|$ and  $|\delta m^2_{32}|$. 

Three experiments are being constructed to look for small values of $\sin^2\theta_{13}$. These are Double Chooz (France), Daya Bay (China) and Reno (South Korea) \cite{Nu2008}.  Double Chooz will start data taking at the end of 2008 with only the far detector with the near detector coming on line in 2009.
The ultimate sensitivity of the Double Chooz experiment is $\sin^22 \theta_{13} = 0.03$ whereas Daya Bay which will start in 2009 has an ultimate sensitivity for  $\sin^22 \theta_{13} < 0.01$. Reno's sensitivity
is comparable to that of Double Chooz.  Neutrino 2012 will be an interesting time for results from these experiments.

The strength of these experiments is that they directly measure $\sin^2\theta_{13}$. However, they
have no sensitivity to the mass hierarchy, the size of $\sin^2 \theta_{23}$ or whether or not CP is violated in the neutrino sector.
 
\begin{figure*}[hbt]
\includegraphics[width=55mm]{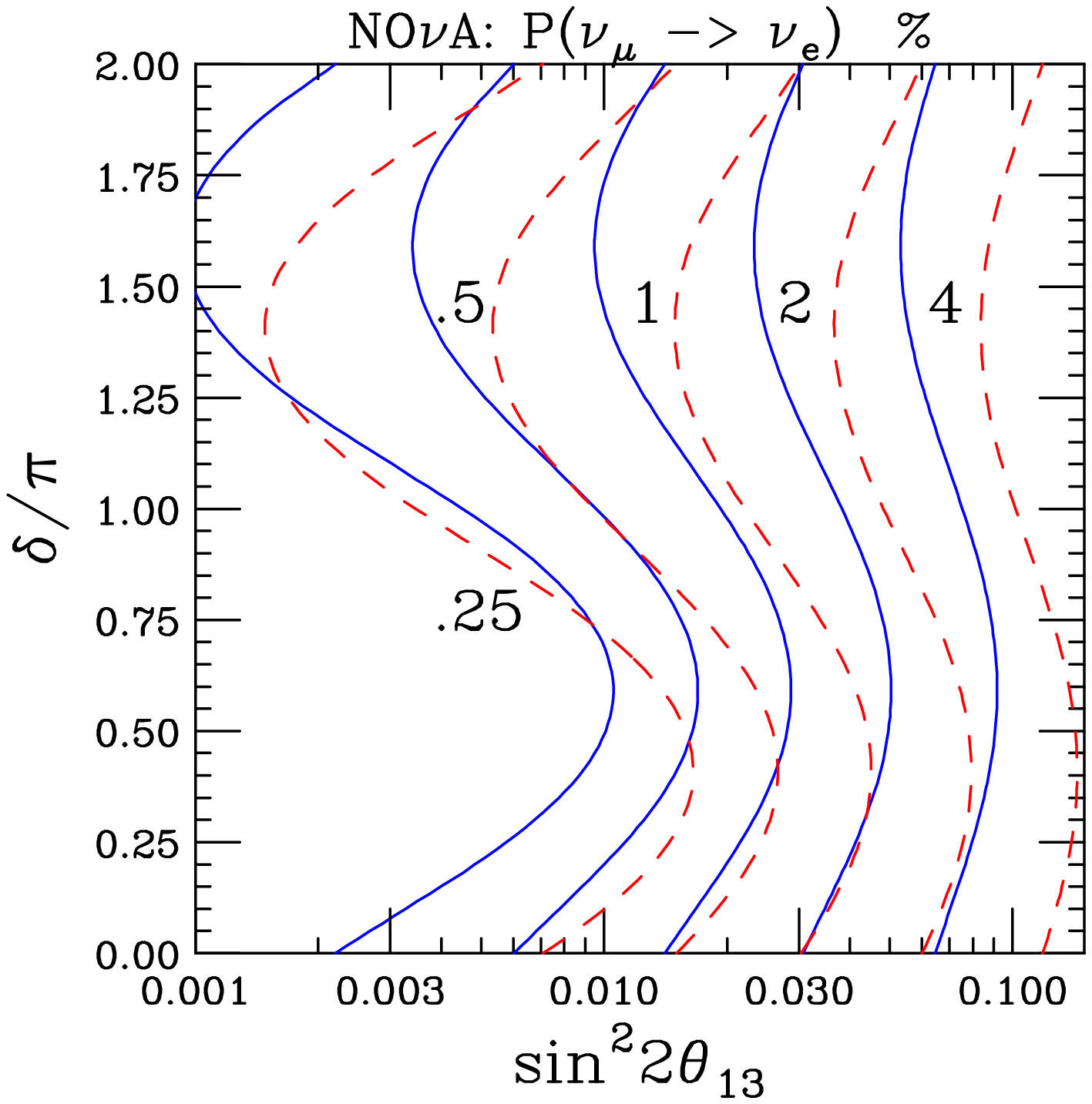}
\includegraphics[width=55mm]{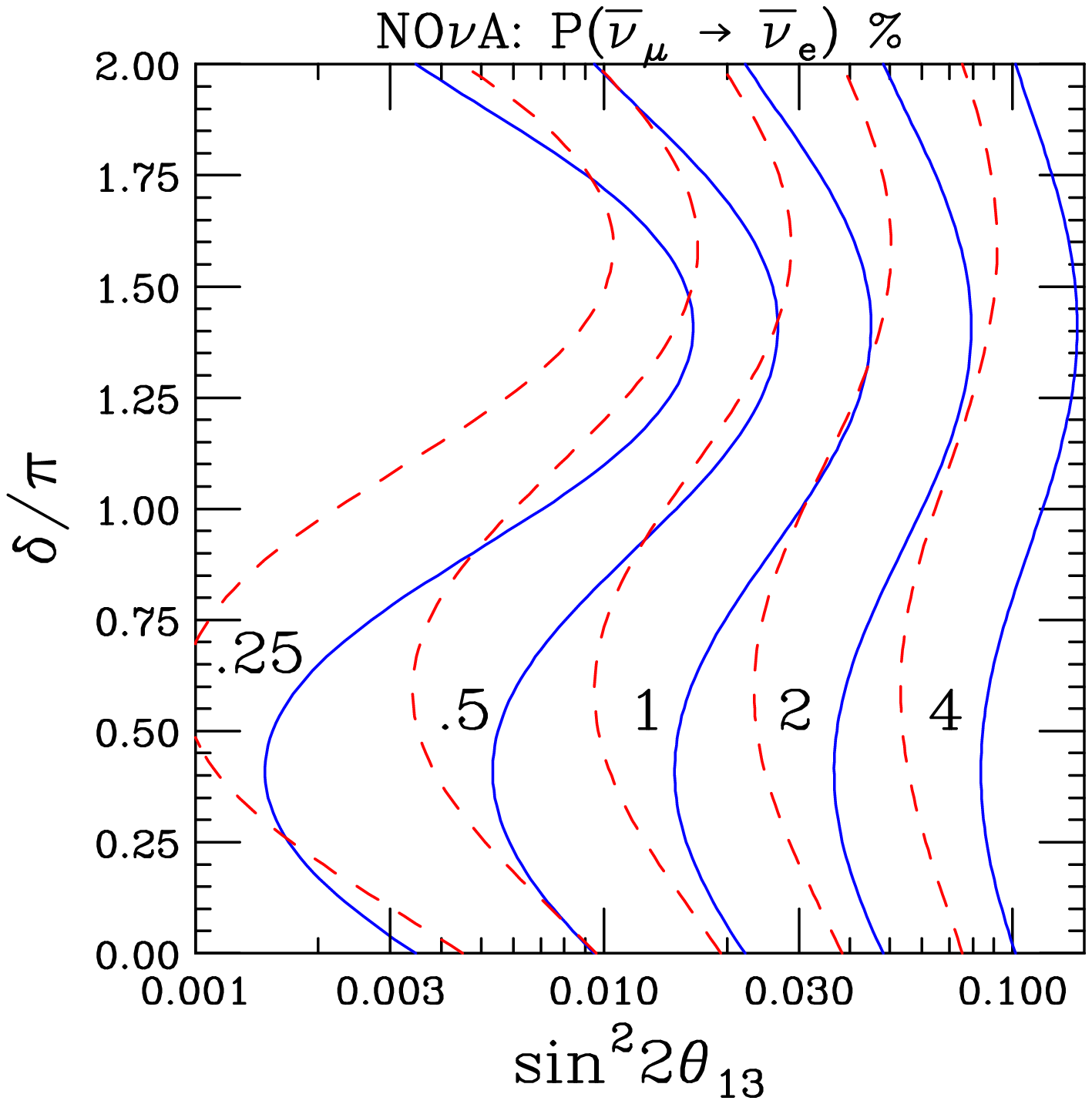}
\includegraphics[width=55mm]{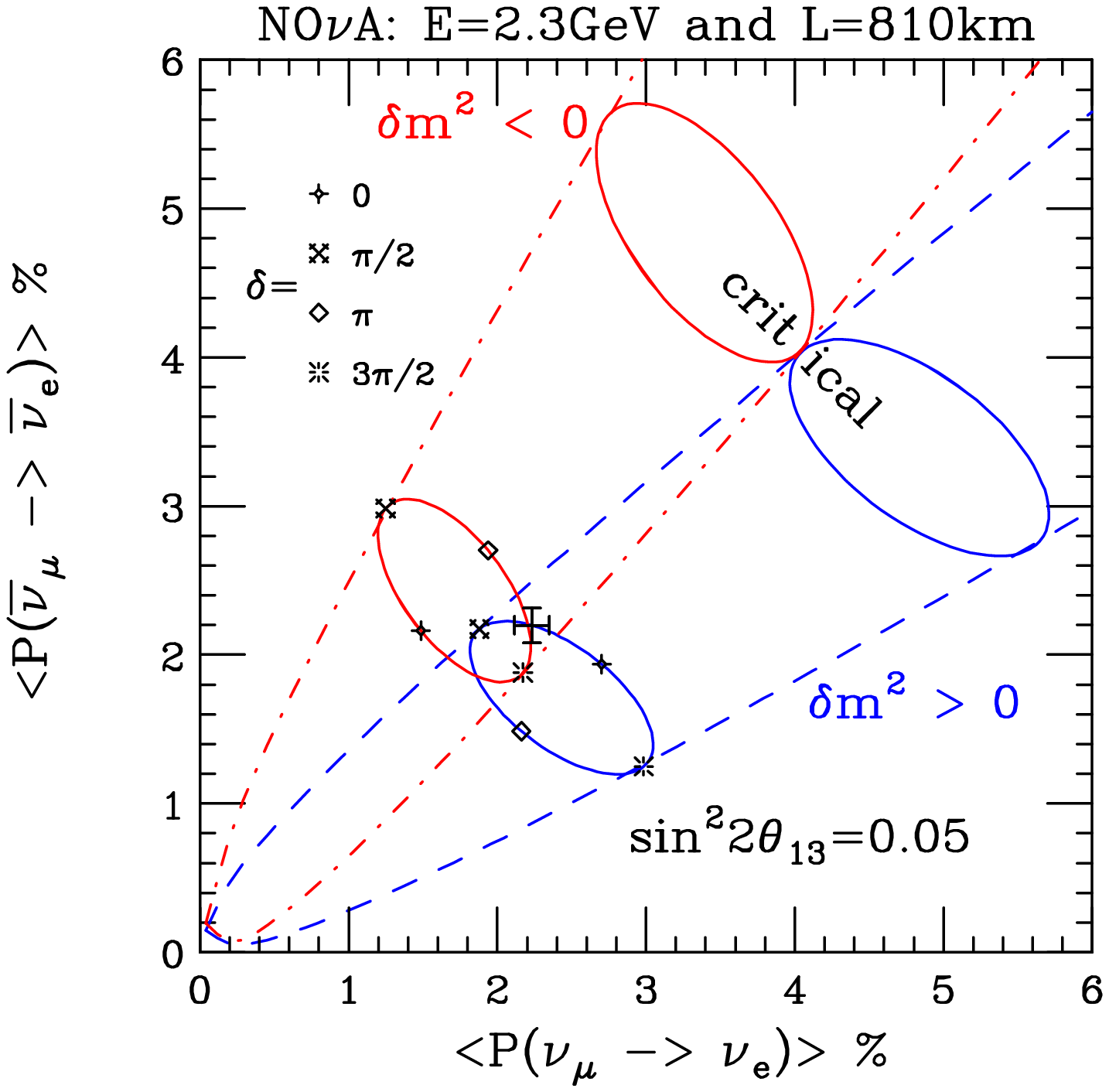}
%\vspace*{-1cm}
\caption{
The left and middle panels are the iso-probability contours for NO$\nu$A as a \% for the neutrino
(left) and anti-neutrino (middle) channels.  The solid (blue) line is for the normal hierarchy whereas the
dashed (red) line is for the inverted hierarchy.
The right panel is the bi-probability plot showing the correlation between the two probabilities.
The matter effects and hence the separation between the hierarchies is 3 times large for T2K than NO$\nu$A primarily
due to the fact NO$\nu$A has three times the baseline as T2K. 
The difference in the matter effect between T2K and NO$\nu$A can be used to
untangle CP violation and the mass hierarchy~\protect\cite{om-sp}.
}
\label{fig:novaprobs}
\end{figure*}

\section{Appearance Channels: {\large \mathversion{bold} $\nu_\mu \rightarrow \nu_e$}}
To address the size of $\sin^2\theta_{13}$, the mass hierarchy, the size of $\sin^2 \theta_{23}$ and whether or not CP is violated in the neutrino sector, the appearance process 
$\nu_\mu \rightarrow \nu_e$ and/or one of its CP and T conjugate processes will need to be measured. That is, in one of following transitions
{
\def \al{\mu}
\def \be{e}
\begin{center}
\hspace*{+0.0cm}\begin{tabular}{lrcll}
& & {CP} & & \\[0.1in]
&{$\nu_\al \to  \nu_\be$} &
$\Longleftrightarrow$ &
%$\red{\bar{\nu}_\al \to  \bar{\nu}_\be}$   & \red{Super-Beams} \\[0.2in]
{$\bar{\nu}_\al \to  \bar{\nu}_\be$}  &  \\[0.2in]
{T} &  $\Updownarrow$ 
 \quad \quad
&\quad \quad{\tiny CPT across diagonals} \quad \quad  & ~~\quad $\Updownarrow$ \quad \quad {T} & \\[0.2in]
&{$\nu_\be \to  \nu_\al$} &
$\Longleftrightarrow$ &
%\blue{$\bar{\nu}_\be \to  \bar{\nu}_\al$} &  \blue{Nu-Fact, $\beta$-Beams} \\
{$\bar{\nu}_\be \to  \bar{\nu}_\al$} &   \\[0.1in]
& & {CP} & & 
\end{tabular}
\end{center}
}
\noindent Processes across the diagonal are related by CPT. The first row will be explored in
very powerful conventional beams, Superbeams, whereas the second row could be explored in Nu-Factories or Beta Beams.

The amplitude for  $\nu_\mu \rightarrow \nu_e$ can be simple written a sum 
of three amplitudes, one associated with each neutrino mass eigenstate, 
{\small \begin{eqnarray}
U_{\mu1}^* e^{-im^2_1L/2E} U_{e1} 
+  U_{\mu2}^* e^{-im^2_2L/2E} U_{e2} 
+  U_{\mu3}^* e^{-im^2_3L/2E} U_{e3}. \nonumber
\end{eqnarray}}
The first term can be eliminated using the unitarity of the MNS matrix and thus the appearance probability can be written 
as follows\cite{Cervera:2000kp}
{%\mathversion{bold}
\begin{eqnarray}
P(\nu_\mu \rightarrow \nu_e) &=& | ~{2U^*_{\mu 3}U_{e3} \sin \Delta_{31} e^{-i\Delta_{32}}} \nonumber \\  & & \quad \quad  \quad \quad + {2 U^*_{\mu2}U_{e2}\sin \Delta_{21}}~|^2 \nonumber \\
& \approx & \left|{\sqrt{P_{atm}}}e^{-i(\Delta_{32}+\delta)} + {\sqrt{P_{sol}}}\right|^{~2}.
\label{pme}
\end{eqnarray}
}
As the notation suggests the amplitude $\sqrt{P_{atm}} $ only depends on $\delta m^2_{31}$ and  $\sqrt{P_{sol}} $  only depends on $\delta m^2_{21}$.
For propagation in the matter, these amplitudes are simple given by 
\begin{eqnarray}
\sqrt{P_{atm}}  & = & \sin\theta_{23} ~\sin 2\theta_{13} ~\frac{\sin(\Delta_{31} - aL) }{( \Delta_{31} - aL)}~\Delta_{31}\nonumber \\
\sqrt{P_{sol}} & = &  ~ \cos\theta_{23} ~ \sin 2 \theta_{12}  
~\frac{\sin ( aL)}{(aL)}~\Delta_{21}.
\end{eqnarray}
\begin{figure}[b]
\begin{center}
\includegraphics[width=40mm]{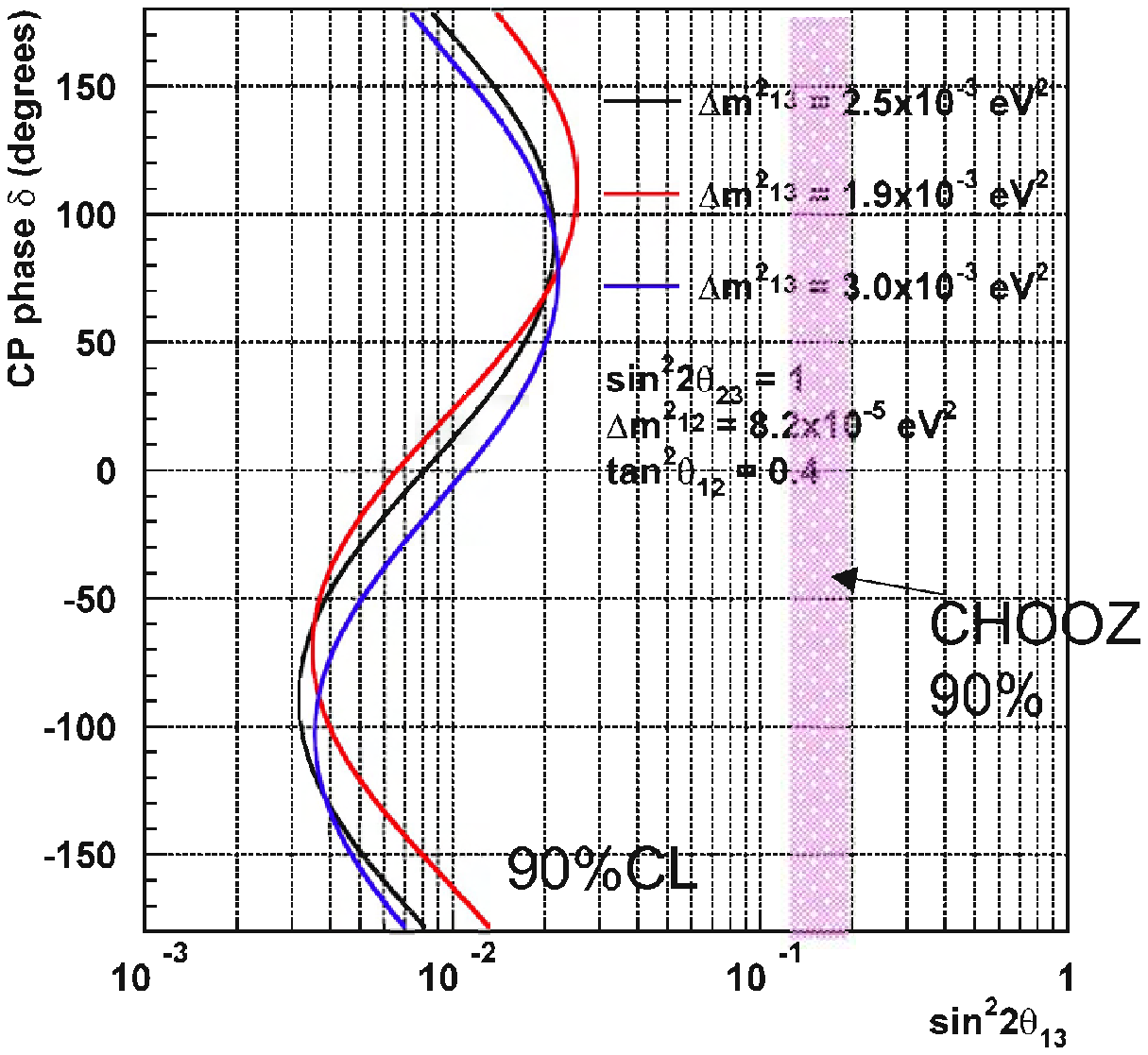}
\includegraphics[width=40mm]{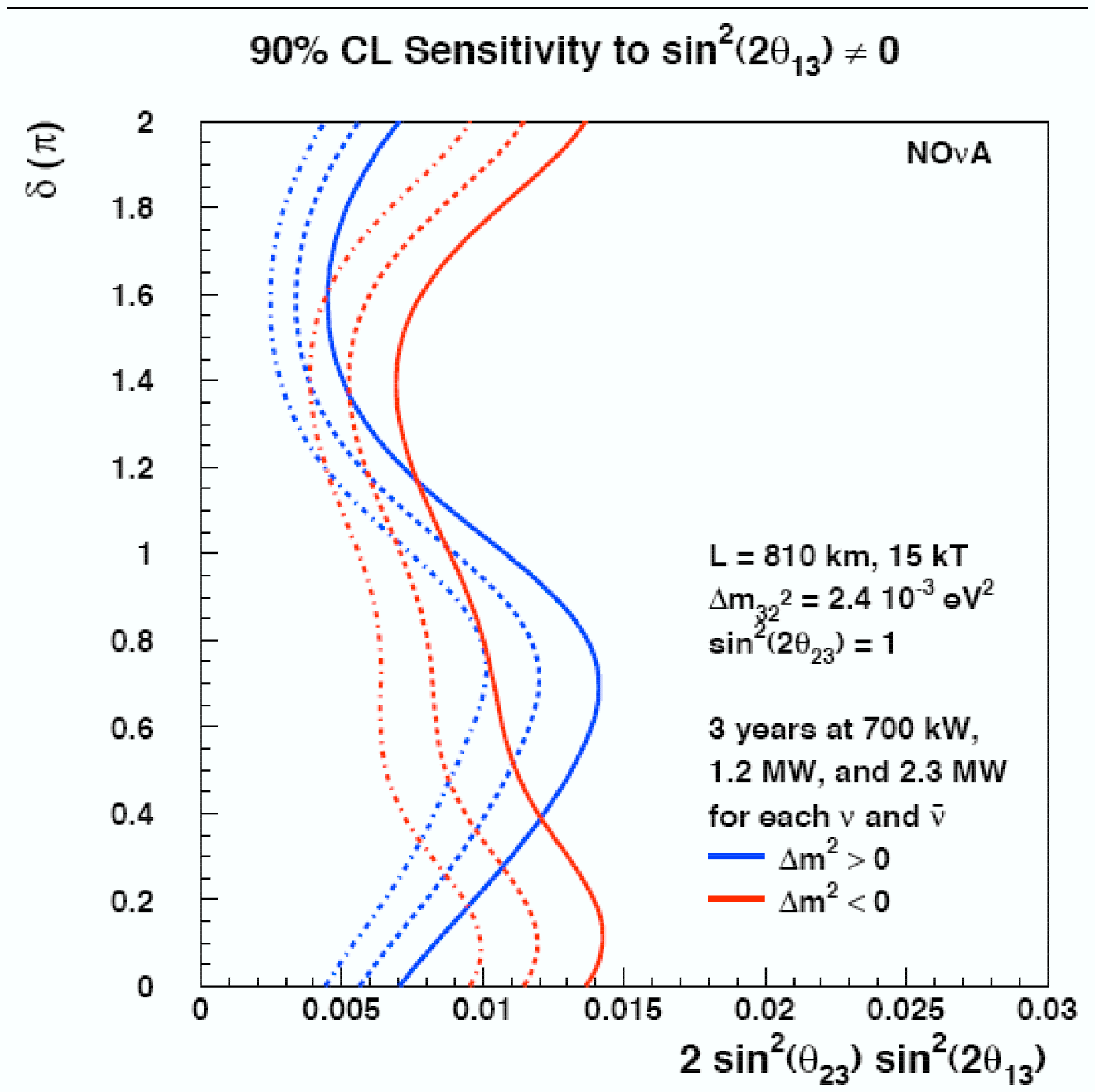}
\end{center}
\caption{Left panel: The 90\% sensitivity to $\sin^22\theta_{13}$ for T2K for 5 years of neutrino running. Right panel: The 90\% sensitivity to $\sin^22\theta_{13}$ for NO$\nu$A
%~\protect\cite{nova} 
assuming 3 years of running time for neutrinos and anti-neutrinos. The blue (red) curves is for the normal (inverted) hierarchy. The three lines
from right to left are for 0.7, 1,2 and 2.3 MW of protons on target respectively. These curves  correspond to $P(\nu_\mu \rightarrow \nu_e)$ and  $P(\bar{\nu}_\mu \rightarrow \bar{\nu}_e)$ at 
the sub 1\% level.}
\label{fig:sens}
\end{figure}
The matter potential is given by 
$a=G_F N_e/\sqrt{2} \approx (4000~km)^{-1}$ and the sign of  $\Delta_{31}$ (and $\Delta_{32}$)
determines the hierarchy; normal   $\Delta_{31}>0$ whereas inverted $\Delta_{31}<0$.  When $a$ is set to zero one recovers the vacuum result. 

\begin{figure}[b]
\begin{center}
\includegraphics[width=40mm]{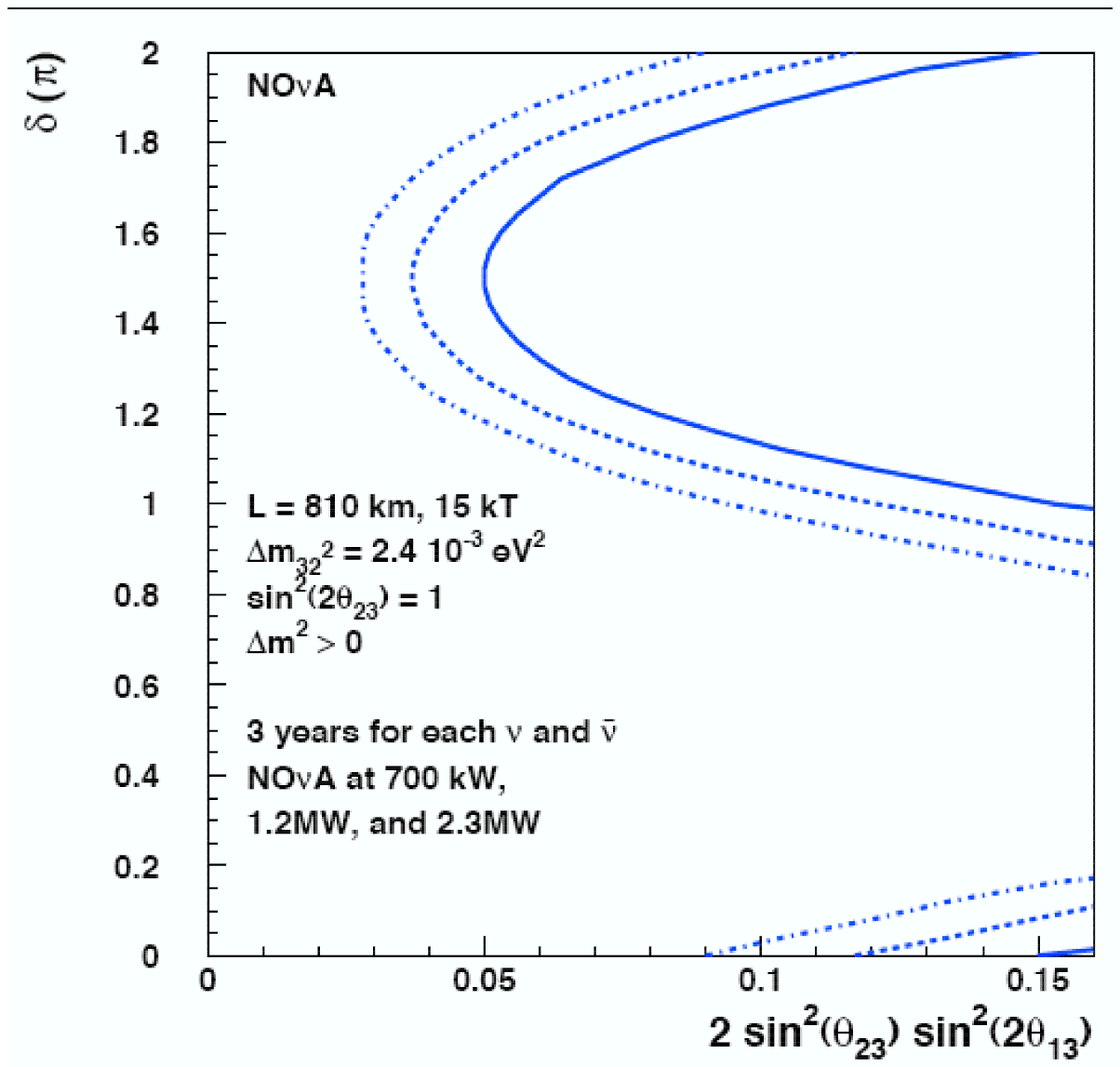}
\includegraphics[width=40mm]{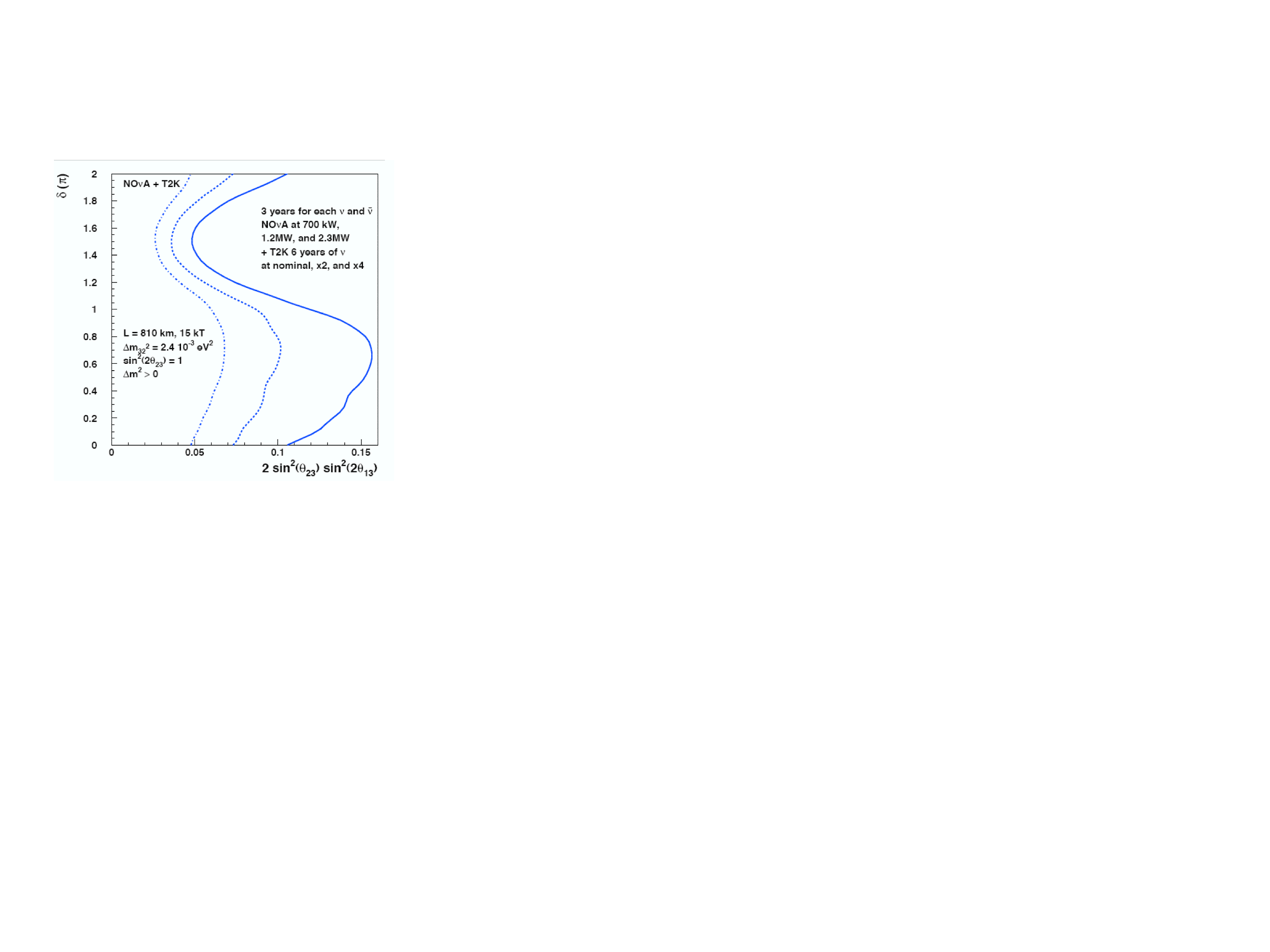}
\end{center}
\caption{The left panel shows the parameters in $\sin^2 2 \theta_{13}$ v $\delta$ plane that NO$\nu$A
%~\protect\cite{nova}  
determines the hierarchy assuming it is normal.
The three lines
from right to left are for 0.7, 1,2 and 2.3 MW of protons on target respectively. The right panel shows the enhancement in the sensitivity when combined with T2K neutrino running only. For the inverted hierarchy the curves are flipped about $\delta=\pi$.}
\label{fig:hier}
\end{figure}

\begin{figure*}[t]
\includegraphics[width=170mm]{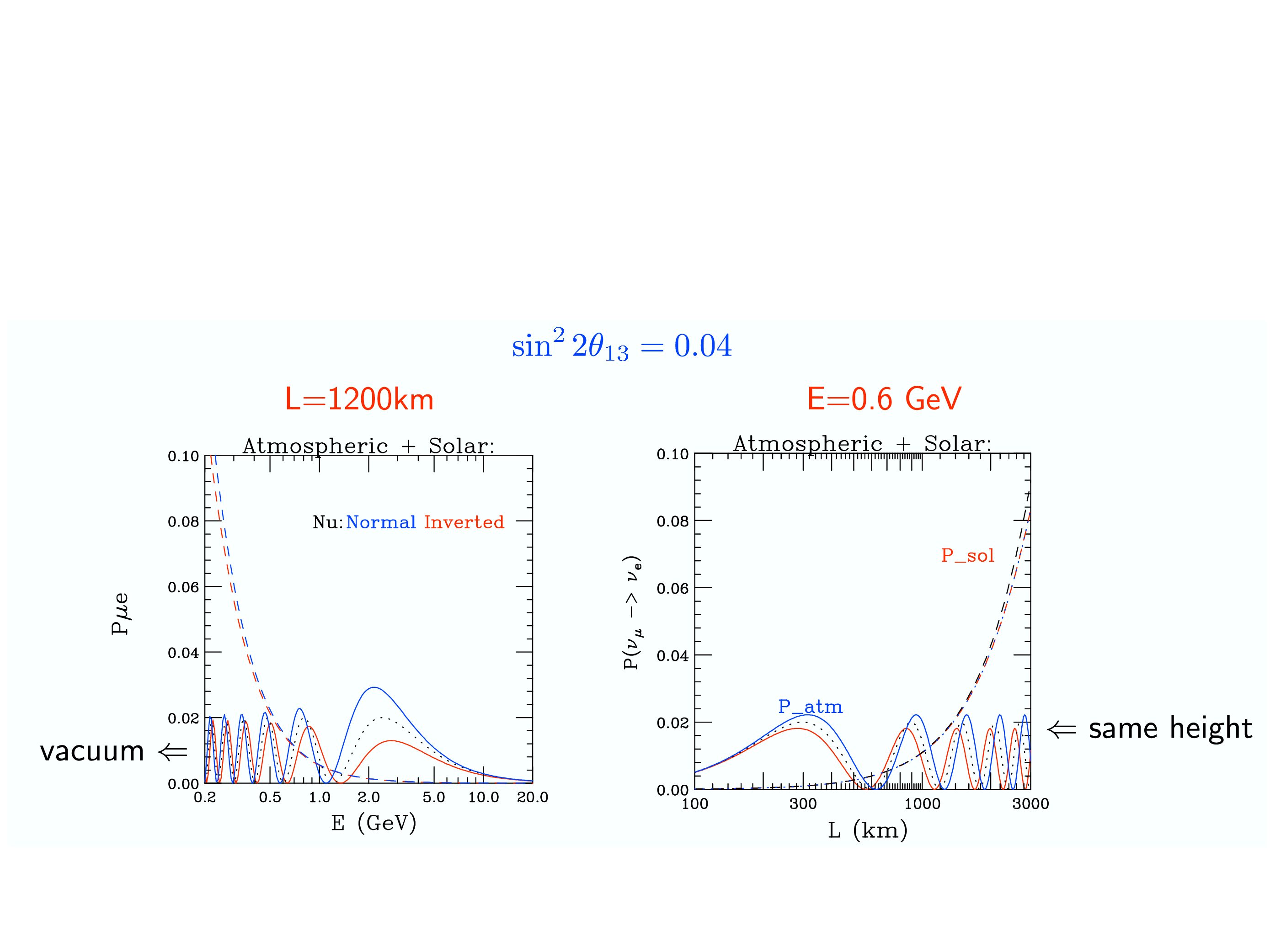}
%\vspace*{-1cm}
\caption{The left panel shows the effects of matter on $P_{atm}$ and $P_{sol}$ holding the baseline
fixed and varying the energy like the Fermilab to DUSEL proposal.  The right panel is the corresponding figure holding the energy fixed and varying the baseline like for the T2KK proposal.
Clearly the effects of matter are very different for these two ways of getting to the second oscillation maximum. This figure was adapted from Ref.~\protect \cite{NPV}.}
\label{fig:beyond}
\end{figure*}

For anti-neutrinos $a \rightarrow -a$ and $\delta \rightarrow -\delta$. Thus the phase between 
 $\sqrt{P_{atm}} $ and $\sqrt{P_{sol}} $ changes from $(\Delta_{32}+\delta)$  to
 $(\Delta_{32}-\delta)$. This changes the interference
 term from 
 {\small \begin{eqnarray}
 2\sqrt{P_{atm}} \sqrt{P_{sol}} \cos(\Delta_{32}+\delta) &  \Rightarrow & 2\sqrt{P_{atm}} \sqrt{P_{sol}} \cos(\Delta_{32}-\delta). \nonumber
\end{eqnarray}}
Expanding  $\cos(\Delta_{32}\pm\delta)$, one has a CP conserving part
$2\sqrt{P_{atm}} \sqrt{P_{sol}} \cos\Delta_{32}\cos\delta$ and the CP violating part
\begin{equation}
\mp 2\sqrt{P_{atm}} \sqrt{P_{sol}} \sin\Delta_{32}\sin\delta. 
\end{equation}
Therefore CP violation is maximum when  $\Delta_{32} = (2n+1)\frac{\pi}{2}$ and grows as 
n grows.
Notice also, that for this term to be non-zero the kinematical phase $\Delta_{32}$ cannot be  $n\pi$. 
This is the neutrino counter part to the non-zero strong phase requirement for CP violation in the quark sector.

The asymmetry between $P(\nu_\mu \rightarrow \nu_e)$ and  $P(\bar{\nu}_\mu \rightarrow \bar{\nu}_e)$ 
is a maximum when  $\sqrt{P_{atm}} =\sqrt{P_{sol}} $. At the first oscillation maximum, $\Delta_{31} = \pi/2$, this occurs when $\sin^2 2\theta_{13} =0.002$  in vacuum.  For values of $\sin^2 2\theta_{13} < 0.002$ the oscillation probabilities are dominated by $P_{sol}$ and thus observing the effects of non-zero $\sin^2 2\theta_{13}$ become increasing more challenging.

Fig. \ref{fig:t2kprobs} and Fig. \ref{fig:novaprobs} give the iso-probability contours for the T2K and NO$\nu$A experiments, see \cite{NPV}.  The third panel in these figures shows the allowed region in neutrino and anti-neutrino bi-probability plane for these experiments.  For NO$\nu$A these allowed regions are significantly separated at large values of $\sin^2 \theta_{13}$. 

Fig. \ref{fig:sens} shows the sensitivity for non-zero $\sin^2 2 \theta_{13}$ for the T2K and NO$\nu$A experiments, see \cite{Nu2008}. Whereas Fig.\ref{fig:hier} shows the region in the $\sin^2 2 \theta_{13}$ v $\delta$ plane for which the hirarchy is determined for NO$\nu$A and the combination of NO$\nu$A with T2K.

Beyond T2K and NO$\nu$A there are a number of proposals  to explore the oscillation probability,
$P(\nu_\mu \rightarrow \nu_e)$, at the second oscillation maximum.  One of these, T2KK, consists of building a second very large detector in Korea so that it is down stream from a similar new large detector
at Kamioka, see \cite{Nu2008}.  Another proposal is to build a new neutrino beamline at Fermilab to send a broad band
neutrino beam to a new very large detector at DUSEL in the Homestake mine, see \cite{Nu2008}.  Here the detector could either be a large version of a water Cerenkov like SuperK or be a very large liquid Argon detector
if this is feasible.  The liquid Argon detector has better $\pi^0$ rejection than the water Cerenkov detectors and also has higher sensitivity to the proton decay channel $p \rightarrow K^+ + \nu$.
The matter effects for these two proposals are quite different.  T2KK experiment gets to the second oscillation peak by using the same energy but three times the baseline whereas Fermilab to DUSEL
uses the same baseline but the energy of the neutrinos at the second oscillation peak is one third
that of the first oscillation peak.  Fig. \ref{fig:beyond} shows the difference in the effects of matter  on the $P_{atm}$ by varying the energy (fixed baseline) and varying the baseline (fixed energy).  Thus the 
difference in the matter effect on the full oscillation probability makes these two proposals complementary and can be used to untangle the effects of matter and CP violation on the oscillation
probabilities and thus determining the neutrino mass hierarchy and whether or not CP is violated in the neutrino sector.

\bigskip % extra skip inserted

\end{document}